# Development of a Transit-Time Ultrasonic Flow Measurement System for Partially Filled Pipes: Incorporating Flow Profile Correction Factor and Real-Time Clogging Detection

Mohammadhadi Mesmarian, Mohammad Mahdi Kharidar, Hossein Nejat Pishkenari*

***Abstract*—Flow measurement in partially filled pipes presents greater complexity compared to fully filled systems, primarily due to the complex velocity distribution within the cross-section, which is a key source of measurement inaccuracy. To address this challenge, an ultrasonic flow meter was designed and developed, capable of simultaneously measuring both flow velocity and fluid level. To improve measurement accuracy, a flow profile correction factor (FPCF) was derived based on the velocity distribution characteristics and applied to the raw flow meter output. A dedicated open-channel flow loop incorporating a 250 mm diameter pipe was constructed to test and calibrate the system under controlled conditions. Flow rates in the loop varied from 2 to 6 liters per second. The accuracy of the flow meter was evaluated using the Flow-Weighted Mean Error (FWME) metric. Experimental results showed that applying the FPCF significantly improved accuracy, reducing the maximum flow measurement error from 8.51% to 2.44%. Furthermore, calibration led to a substantial decrease in FWME from 1.78% to 0.08%, confirming the effectiveness of the proposed methodology. The flow meter was also subjected to clogging scenarios by artificially obstructing the flow. Under these conditions, the device was able to reliably measure the flow and successfully detected the clogging, triggering an alarm to the operator to take necessary action.**

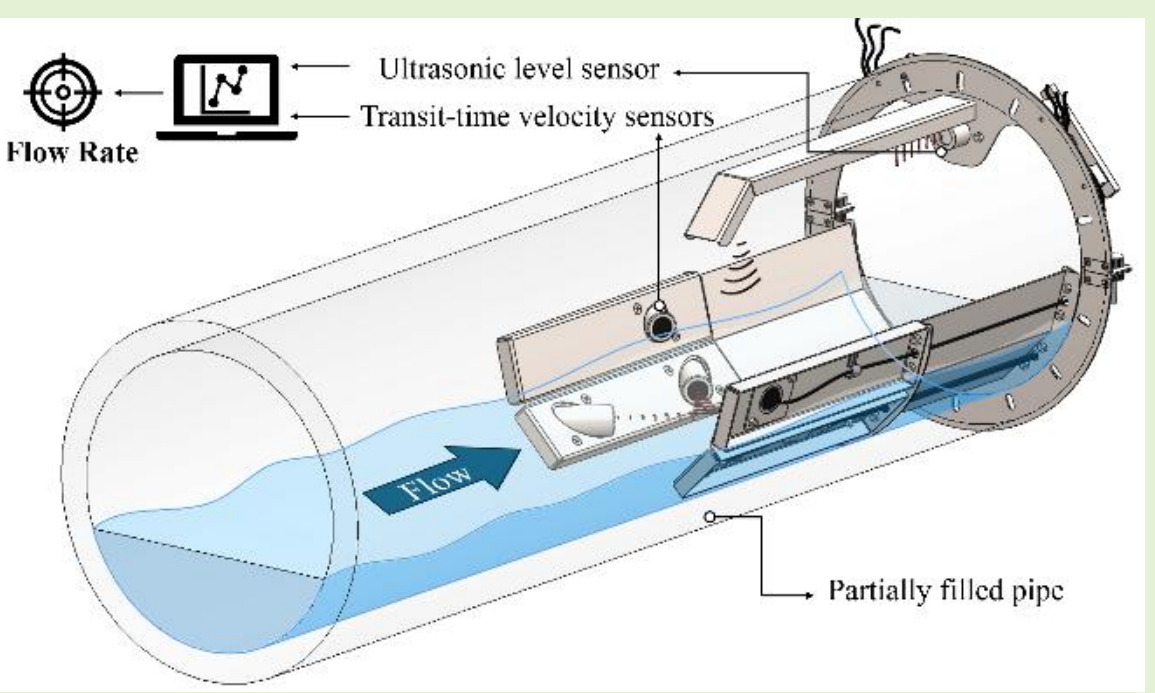

## I. Introduction

FLOW measurement is crucial in many industrial processes, such as petroleum, chemical, energy, and wastewater management [1]. While fully filled pipes are commonly used for fluid transport, many applications involve partially filled pipes, including petrochemical processes, food and personal care industries, and sewer systems [2], [3]. Flow measurement under these conditions is more challenging due to the presence of a free surface, which introduces water level fluctuations, surface tension effects, and secondary currents [4], [5]. Additionally, the velocity distribution varies with flow depth, with slower velocities near the bottom and maximum velocity occurring near or just below the free surface, complicating accurate measurement with traditional techniques [6].

Andrew Godly reviewed a wide range of methods for measuring flow in partially filled closed conduits, summarizing traditional techniques such as volumetric measurements, tracer-dilution methods, and gauging structures, as well as advanced approaches including ultrasonic, electromagnetic, and non-contact technologies [7]. Traditional devices, such as propeller-type current meters combined with pressure transducers in siphonic roof drainage systems, can introduce turbulence, suffer from calibration issues, and exhibit limited sensitivity at low flows [8]. Pressure transducers also face challenges from debris accumulation and potential diaphragm damage. An alternative approach using a U-shaped communication pipe with a high-precision internal pressure sensor has been proposed, employing a liquid-level-to-flow function and CFD simulations to improve accuracy, achieving measurement errors below 2% for liquid levels below 10% of the pipe diameter [9].

Electromagnetic flow meters are widely used due to their high accuracy and broad measurement range. A specialized electromagnetic flow meter for partially filled pipes, designed based on non-full flow velocity distributions, was developed and tested in a laboratory calibration setup, achieving experimental errors below 1.3% [10]. Limitations include applicability only to conductive liquids, high weight, lack of modularity, and restricted pipe diameter ranges.

Vision-based techniques have also been explored for partially filled pipes and open channels. Nguyen et al. developed an in-situ sewer monitoring system using video images to estimate water levels, and Wook Ji et al. combined image processing with deep learning to measure flow rates in sewage pipes [11], [12]. These methods rely on detecting fluid boundaries and applying hydraulic equations such as Manning's formula. However, their performance is sensitive to lighting conditions and high-tech hardware requirements, limiting suitability for continuous, real-time monitoring.

Ultrasonic flow measurement is widely used in industrial applications, with three main methods: Doppler, cross-correlation, and transit-time. All can be applied to partially filled pipes and open channels, though transit-time sensors are preferred for their modularity, high accuracy, fast response, low pressure loss, and minimal maintenance [7], [13], [14]. Munasinghe and Paul employed a transit-time ultrasonic flow meter combined with a four-channel capacitance sensor to improve accuracy in partially filled pipes, achieving a 92% reduction in mean error compared to conventional ultrasonic meters at low water heights [15].

Several studies have further addressed flow measurement in rectangular or complex channel cross-sections, incorporating velocity distributions via ultrasonic sensors [16]-[21]. For example, Bonakdari and Zinatizadeh used CFD simulations to derive correction coefficients linking Doppler-measured velocities to mean flow velocities in an egg-shaped sewer channel [19]. While effective for their specific setup, these results are not directly generalizable to other channel geometries or circular pipes.

Despite recent advancements, highly accurate, real-time methods for measuring flow in partially filled circular pipes using transit-time ultrasonic sensors remain limited, particularly those accounting for the velocity distribution. In this study, we develop an ultrasonic-based flow measurement system that simultaneously captures fluid velocity and water level. The key innovation is the incorporation of a flow profile correction factor, which compensates for cross-sectional velocity variations and can be generalized to different pipe diameters and sensor configurations, enhancing the method's robustness and applicability.

## II. METHODOLOGY

The flow rate $Q$ is given by the equation $Q = A\overline{v}$ , where $A$ is the cross-sectional area and $\overline{v}$ is the average velocity across that area. Alternatively, the flow rate can be calculated using the integral form (1).

$$Q = \int_A v(x, y) dA \tag{1}$$

In this research, a numerical model to calculate the normalized velocity distribution in partially filled pipes is developed. Unlike many existing methods, which are constrained by the physical characteristics of specific pipes or channel geometries, the model provides greater generalizability across various pipe diameters and flow conditions.

The following section discusses the methodology for measuring flow velocity and cross-sectional area with ultrasonic sensors. It describes the final step of calculating the flow rate using an appropriate flow profile correction factor.

### A. The principle of the transit-time ultrasonic velocity measurement

The transit-time ultrasonic velocity measurement provides an accurate method for determining flow velocity in pipes. Ultrasonic waves are generated and received using piezoelectric transducers. In this method, the wave travels once from the upstream transducer to the downstream transducer and once in the reverse direction. The difference in travel time in both directions is used to calculate the flow velocity using Equation (2) where $t_{up}$ is the time for the pulse to travel from upstream to downstream, $t_{down}$ is the time to travel from downstream to upstream, $L$ is the distance between the two transducers, and $\theta$ is the angle between the sound path and the flow direction [22].

$$v_{Line} = L \frac{t_{down} - t_{up}}{2 t_{up} t_{down} \cos\theta} \tag{2}$$

### B. The principle of the ultrasonic level measurement

Determining both the flow velocity and the cross-sectional area is essential for accurately measuring the flow rate. The level sensor helps determine the liquid level in the pipe. Then, based on the pipe's geometry and mathematical calculations, the cross-sectional area of the flow passing through the pipe can be calculated. The cross-sectional area $A$ of the fluid in a partially filled circular pipe can be calculated from the measured flow level $H$ and knowing the pipe diameter $D$ (shown in Fig. 1), using the following formula:

$$A = \frac{\pi R^2 \theta}{2\pi} + \frac{1}{2} R^2 sin(2\pi - \theta) = \frac{D^2}{8}(\theta - \sin\theta) \tag{3}$$

Where:

$$\theta = 2\cos^{-1}(1 - \frac{2H}{D}) \tag{4}$$

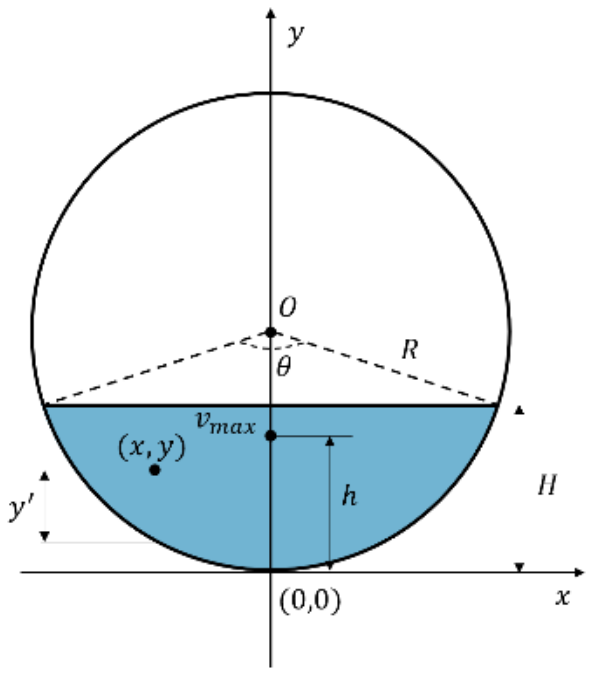

Fig. 1. Parameters of partially filled cross-section

### C. Determining Correct Flow Rate with Flow Profile Correction Factor

In this research, the hybrid technology to measure flow rate is used. This method utilizes transit-time velocity transducers,

which are positioned across the pipe, to measure the flow velocity at a specific section. Additionally, an ultrasonic level sensor is used to measure the cross-sectional area of the flow as a function of level. However, simply multiplying the calculated cross-sectional area by the measured velocity does not provide an accurate flow rate measurement. This is due to the fact that the velocity measured by the transit-time transducers at a chordal path across the pipe is not a good approximation of the average flow velocity. To address this, we investigate a correction factor that accounts for the velocity distribution in a partially filled pipe. This factor, called the Flow Profile Correction Factor (FPCF), is designed to improve flow rate measurement accuracy. The first step in obtaining this value is to extract a velocity distribution in a partially filled circular pipe numerically. Unlike fully filled pipes, there is limited research on the velocity distribution in partially filled circular pipes. Jiang et al. derived a 2D velocity distribution for partially filled circular pipe flow based on the principle of maximum entropy [23]. Although the formulas presented in their research are complex, they are useful. We simplified and adapted these formulas to derive the FPCF for flow rate measurement in partially filled pipes. In Fig. 1. the schematic of partially filled circular pipeline channel is shown. $R$ represents the radius of the pipe, $H$ is the water level, and $h$ ( $0 \leq h < H$ ) denotes the level of maximum velocity point ($v_{max}$). A rectangular coordinate system is defined with its origin at the pipe's inner bottom. The x-axis represents the horizontal distance from the centerline, while the y-axis indicates the vertical level from the x-axis in the upward direction. The coordinates (x, y) correspond to an arbitrary point within the pipe's cross-section. The general expression for the velocity distribution is given by:

$$v = (1-\frac{1}{M})v_{\max} + \frac{v_{\max}}{M}\left(\frac{y'(1-(1-M)^{\frac{q}{q-1}})F(v)}{y} + (1-M)^{\frac{q}{q-1}}\right)^{1-\frac{1}{q}} \qquad (5)$$

In this formulation, two parameters appear: $M$, a dimensionless entropy parameter, and $q$, the non-extensive entropy parameter from Tsallis entropy theory. Both were calibrated experimentally by relating them to the mean velocity, maximum velocity, and the position of maximum velocity at different flow depths [23]. In this work, we adopt the validated values $M = 0.89$ and $q = 1.15$ as reported in [23].

Where $F(v)$ is the cumulative distribution function of the velocity distribution in a partially filled circular pipe as follows:

$$F(v) = 4\left(\left(\frac{y'}{2R}\right)^{s} - \left(\frac{y'}{2R}\right)^{2s}\right)\left(1-\left(\frac{y'}{h'}-1\right)^{2L}\right)^{K}\left(1-\left(\frac{x}{R}\right)^{(\frac{D}{H})}\right) \qquad (6)$$

Where:

$$y' = y - (R - \sqrt{R^2 - x^2})$$

$$s = \frac{\ln 2}{\ln(2R) - \ln h'}$$

$L = 1, K = 1$ for $y' \leq h'$

$L = \dfrac{2h}{H}, K = \dfrac{2(H-h)}{H}$ for $y' > h'$

In the formulas (5) and (6), there are several parameters need to be determined for these formulas to be applicable. For example, $h'$ represents the vertical distance from the maximum velocity to the pipe bed at any vertical cross-section. The formula for calculating the $h'$ requires the calculation of $v_{max}$, which is determined experimentally. To overcome that, we have developed an interpolating function which relates the dimensionless vertical distance of maximum velocity to the dimensionless level based on experimental data presented in [23]. The data corresponds to three dimensionless flow levels: 36.2%, 50%, and 70%. We fitted a third-degree polynomial curve to five data points: three from the experimental data, and two additional points assuming that the pipe is either completely full (100%) or empty (0%). Several parameters are used in the interpolation: $h$ is the vertical distance from the maximum velocity point to the pipe bed at $x = 0$; $h'$ is the vertical distance from the maximum velocity point to the pipe bed at any vertical cross-section; $H$ is the flow level; and $H'$ is the vertical distance from the free surface of the flow to the pipe bed at any vertical cross-section. The expression (interpolating formula) for the vertical distance of the maximum velocity to the pipe bed is:

$$\frac{h}{H} = 1.78\left(\frac{H}{D}\right)^3 - 2.46\left(\frac{H}{D}\right)^2 - 0.18\left(\frac{H}{D}\right) + 1.00 \qquad (7)$$

After deriving the velocity profile equation for a partially filled pipe, we can now proceed to formulate the FPCF as a function of flow level within the pipe. As mentioned earlier, we normalized the velocity equation (5) because the term $v_{max}$ is not directly measurable and determining it requires complex instrumentation. The resulting normalized velocity equation is:

$$\frac{v}{v_{\max}} = 1 - \frac{1}{M} + \frac{1}{M}\left(\frac{y'(1-(1-M)^{\frac{q}{q-1}})F(v)}{y} + (1-M)^{\frac{q}{q-1}}\right)^{1-\frac{1}{q}} \qquad (8)$$

To calculate the mean velocity, integration is performed over the cross-sectional area.

$$\overline{v} = \frac{1}{A}\int_A v(x,y)\,dA \qquad (9)$$

Because the velocity term $v_{max}$ is undefined, the normalized velocity is used for the integration as follows:

$$\hat{v}_A = \frac{1}{A}\int_A \frac{v(x,y)}{v_{\max}}\,dA \qquad (10)$$

Equation (10) gives the normalized mean velocity over the cross section. Similarly, the mean chordal velocity is derived by integrating the normalized velocity along that chord. Equation (11) gives the normalized mean velocity along a horizontal line ($L$) at a given height ($Y$) that is the location of the transit-time velocity sensors in the cross section of the pipe.

$$\hat{v}_L = \frac{1}{L}\int_L \frac{v(x,Y)}{v_{\max}}\,dx \qquad (11)$$

Therefore, the Flow Profile Correction Factor is defined as:

$$FPCF = \frac{\hat{v}_A}{\hat{v}_L} = \frac{\frac{1}{A}\iint \frac{v(x,y)}{v_{\max}}\, dxdy}{\frac{1}{L}\int \frac{v(x,Y)}{v_{\max}}\, dx} \tag{12}$$

Accordingly, the accurate flow rate at each transducer position is calculated using Equation (13), where velocity is measured by the transit-time transducers and the flow level is measured by the ultrasonic level sensor. Equation (14) presents a generalized form for calculating the flow rate when additional transducers are placed at different heights within the cross-section to measure velocity.

$$Q = FPCF(\frac{H}{D}) \times \overline{v}_L \times A(\frac{H}{D}) \tag{13}$$

$$Q = \sum_{i=1}^{n} w_i \times FPCF_i(\frac{H}{D}) \times \overline{v}_{L_i} \times A(\frac{H}{D}) \tag{14}$$

## III. Experimental Section

The following section is divided into two main parts. In the first subsection, the designed and constructed ultrasonic flow meter for measuring flow rates in partially filled pipe is presented and in the second subsection the test setup that was fabricated for test and calibration of the flow meter is introduced.

### A. Design and Construction of the Ultrasonic Flow Meter

To measure the flow rate in partially filled pipes, a flow meter based on ultrasonic technology has been designed and constructed. The device is equipped with six transit-time velocity transducers, which are arranged to form two crossed acoustic paths in the lower section and one acoustic path in the upper section. This configuration is intended to reduce the uncertainty and errors typically associated with single-path measurements [24]-[26]. A level sensor is also integrated to determine the flow level and, subsequently, the cross-sectional area of the flow. In the present study, however, only the acoustic paths located in the lower section are utilized, as the experimental conditions have been set to measure low flow rates in all tests. In Fig. 2, the pictures of the designed and constructed ultrasonic flow meter are shown.

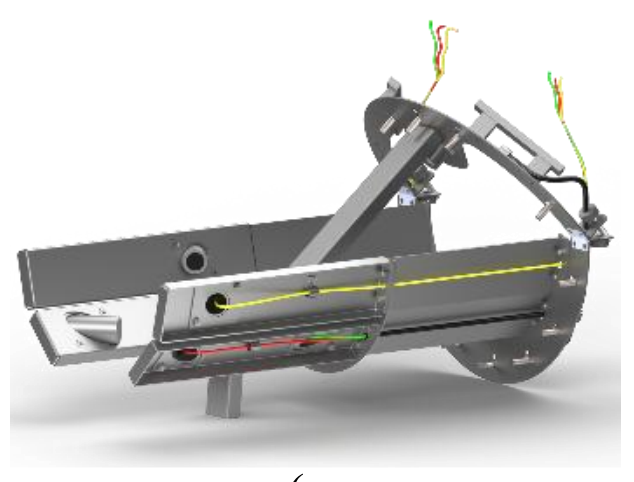

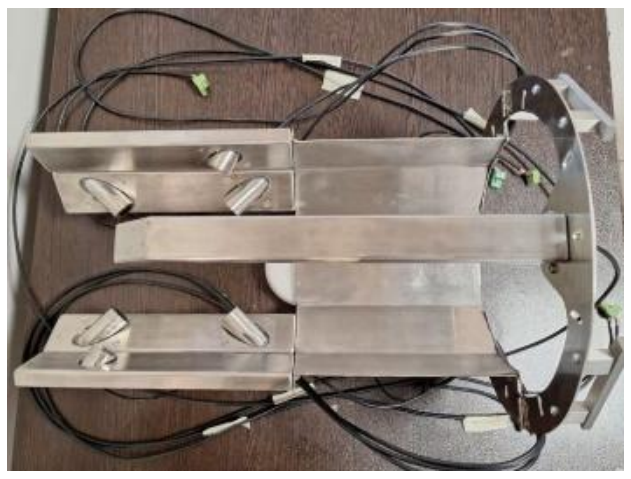

(a (b)

Fig. 2. Ultrasonic flow meter; (a) Designed. (b) Constructed.

### B. Open Channel Flow Loop Test Setup

To facilitate the test and evaluation process of the ultrasonic flow meter, a custom flow loop test system was developed and constructed for open-channel (partially filled pipe) flow measurement and calibration in the Ultrasonic Laboratory of the Mechanical Engineering Department at Sharif University of Technology. The system operates by recirculating water from a storage tank through a piping network using a water pump. The upstream section of the pump was equipped with a 2-inch pipe and an Endress+Hauser Promass 80S Coriolis flow meter, which acted as the reference flow meter. This reference (master) flow meter delivers real-time flow rate measurements with an accuracy of better than 0.1%. Downstream of the pump, the system transitions to 4-inch pipes, which connect to a 10-inch corrugated pipe via two reducer-expanders. The fabricated ultrasonic flow meter was installed at the downstream end of the corrugated pipe. Water flows through this section and cascades into the storage tank, effectively simulating open-channel flow conditions. The corrugated pipe, approximately 2.5 meters in length, provides sufficient distance for the flow to fully develop before measurement. The schematic design of the flow loop test setup and the photograph of the constructed setup are presented in Fig. 3.

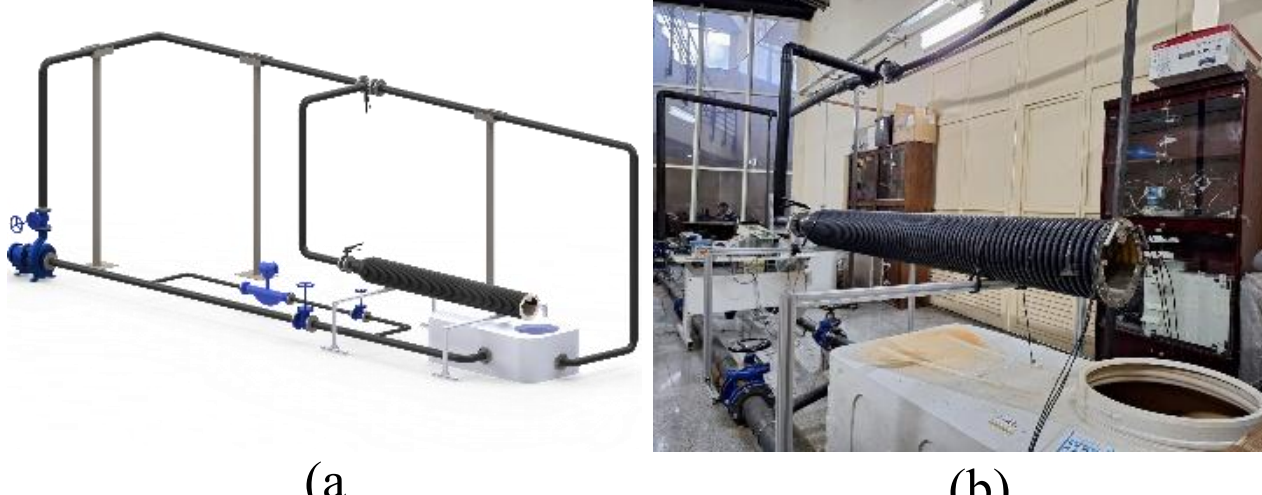

(a (b)

Fig. 3. Open channel flow loop test; (a) Designed. (b) Constructed.

## IV. Numerical Analysis Section

In this section, a methodology is presented for deriving a formula for the flow profile correction factor, specifically adapted to the constructed ultrasonic flow meter and a targeted flow rate range. Initially, the normalized velocity profile is visualized for a partially filled 250 mm diameter pipe. Subsequently, a tailored formula for computing the flow profile correction factor is developed for the flow meter.

### A. Normalized velocity profile in the partially filled pipe

As previously outlined, the velocity profile is normalized due to the inaccessibility of direct measurement of the maximum velocity, which would otherwise require advanced and complex instrumentation. To investigate the normalized velocity distribution, a Python-based computational tool was developed to generate flow velocity profile within a circular conduit of 250 mm diameter. The resulting velocity profiles were generated for three representative flow levels: at half the pipe diameter ($H$ = 125 mm), below half ($H$ = 100 mm), and above half ($H$ = 150 mm) as shown in Fig. 4.

A Python-based computational tool was implemented to numerically evaluate the velocity distribution equations, and the generated profiles showed close agreement with the benchmark distributions reported in [23].

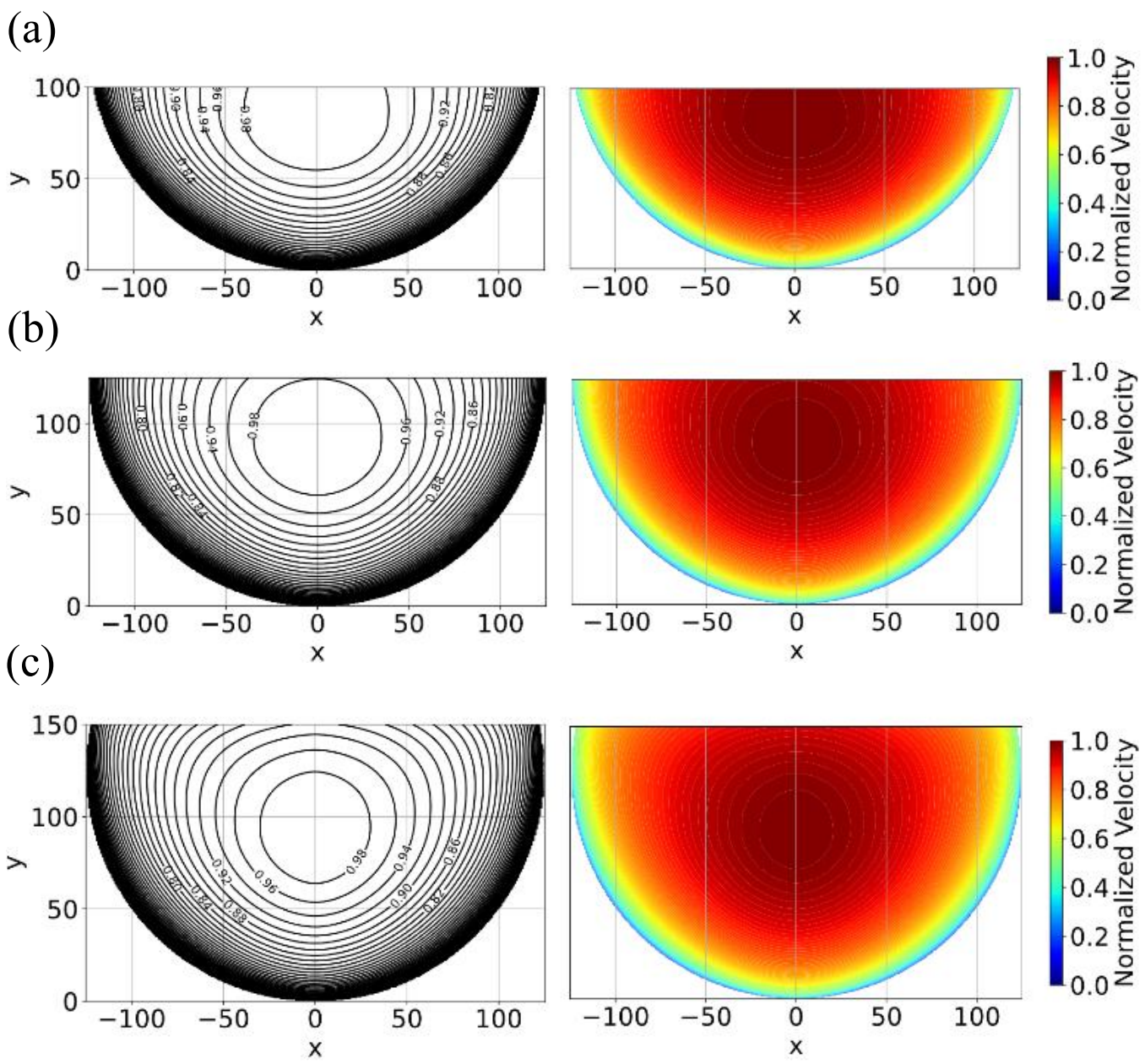

Fig 4. Normalized velocity profiles in a pipe with a diameter of 250 mm: (a) flow level below half-full (H = 100 mm); (b) half-full flow level (H = 125 mm); and (c) flow level exceeding half-full (H = 150 mm).

## B. Deriving the Flow Profile Correction Factor

Owing to the mathematical complexity of Equation (12), an analytical solution of the integrals is not tractable. Therefore, numerical integration techniques are employed to evaluate the expressions for a range of flow levels, $H$. The FPCF is computed at discrete intervals of 10 mm over the level range from $H = 50$ mm to $H = 250$ mm. To obtain a continuous functional representation of the FPCF, a sixth-degree polynomial is fitted to the numerically obtained data. The resulting polynomial expression characterizes the FPCF as a function of the flow level ($H$), and is given by

$$FPCF(H) = 4.22\times10^{-14} H^{6} - 1.35\times10^{-11} H^{5} - 1.96\times10^{-9} H^{4} + 1.24\times10^{-6} H^{3} - 1.81\times10^{-4} H^{2} + 1.24\times10^{-2} H + 6.03\times10^{-1} \quad (15)$$

Although piecewise polynomial spline fitting can provide a closer match to the experimental data, it introduces a segmented relationship between water level and the flow profile correction factor. This approach presents two main drawbacks. First, its implementation on microcontrollers poses challenges due to constraints on processing speed and memory. Second, instead of yielding a single, unified analytical expression, the result is a piecewise formula with increased complexity. From both design and analytical standpoints, a simple, single-expression formula is far more desirable. Therefore, a sixth-degree polynomial was selected to fit the data points, offering a practical compromise between accuracy and simplicity. This form facilitates easier integration into both system analysis and embedded implementation. Equation (15) is the polynomial approximation obtained by numerically evaluating Equation (12) across the studied flow levels and fitting the results. Equation (15) is valid for flow levels in the range 50 mm $\leq H \leq$ 250 mm. This constraint is due to the lack of velocity data for levels below 50 mm, corresponding to the region beneath the measurement domain of the transit-time ultrasonic sensors. Consequently, the FPCF cannot be defined for flow levels less than 50 mm. Fig. 5 illustrates the curve fitted to the data points derived from the numerical solution of Equation (12).

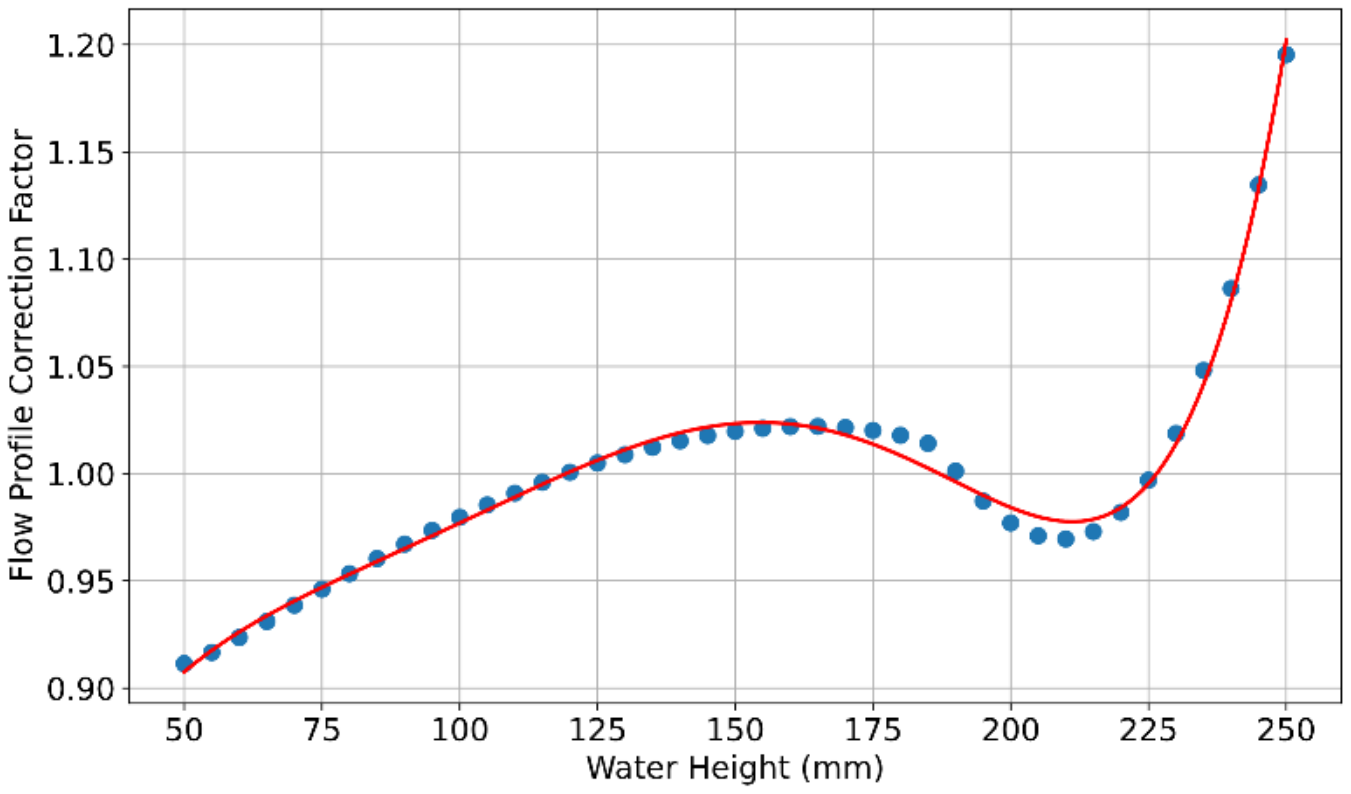

Fig. 5. Sixth-degree polynomial fitted curve for driving flow profile correction factor formula

## V. Results and Discussion

This section presents the experimental results obtained from a series of tests conducted under a range of controlled operating conditions. The evaluation begins with a baseline assessment of the constructed ultrasonic flow meter's performance under nominal conditions. For this purpose, the experimental setup was configured to produce flow rates between 2 and 6 liters per second. This initial phase, referred to as the baseline test, serves to establish a benchmark for evaluating the intrinsic accuracy and effectiveness of the flow meter. During this test, three sets of data are compared: the raw (unprocessed) measurements, measurements corrected using the FPCF, and the fully calibrated outputs. Subsequent to the baseline assessment, the repeatability of the flow meter was investigated at each flow rate to evaluate the consistency and reliability of the measurements. To assess the flow regime in our experiments, the Reynolds number was calculated according to [3]. We obtained $Re_H \approx 3.0\times10^4$ at the lowest tested flow condition ($Q$ = 2 L/s, $H$ = 65 mm)) and $Re_H \approx 7.0\times10^4$ at the highest tested flow condition ($Q$ = 6 L/s, $H$ = 100 mm). Both values are well above the laminar-to-turbulent transition threshold, confirming that all experiments were carried out in the fully turbulent regime. It should be noted that the experimental validation conducted in this study was limited to flow rates in the range of 2–6 L/s, which correspond to fully turbulent conditions in our setup. The applicability of the proposed method in laminar or transitional flows, as well as under extremely high-flow conditions with strong surface fluctuations and air entrainment, may require additional verification. Moreover, hydraulic conditions such as pipe slope and internal roughness can influence the velocity distribution and may necessitate recalibration of the FPCF. Investigating the performance of the system under different pipe slopes and hydraulic conditions is one of our planned future research directions. Finally, the device's response to partial blockage and clogging was investigated. In these cases, the primary objective shifts from precise flow measurement to reliable detection and signaling of abnormal flow conditions. To simulate partial obstructions, two transparent acrylic weirs of varying heights were fabricated and positioned downstream of the measurement section. These tests enabled the evaluation of the system's capability to detect obstructions and generate appropriate alerts for operational awareness and system integrity.

### A. Baseline Test of the Ultrasonic Flow Meter

The performance of the constructed ultrasonic flow meter was evaluated using the open-channel flow loop test setup across five discrete nominal flow rates: 2, 3, 4, 5, and 6 liters per second. Each target flow rate was established through valve adjustments and verified using the Coriolis reference flow meter to ensure measurement accuracy. Fig. 6 presents the aggregated results for each flow condition, providing the complete dataset employed in the derivation of the FPCF.

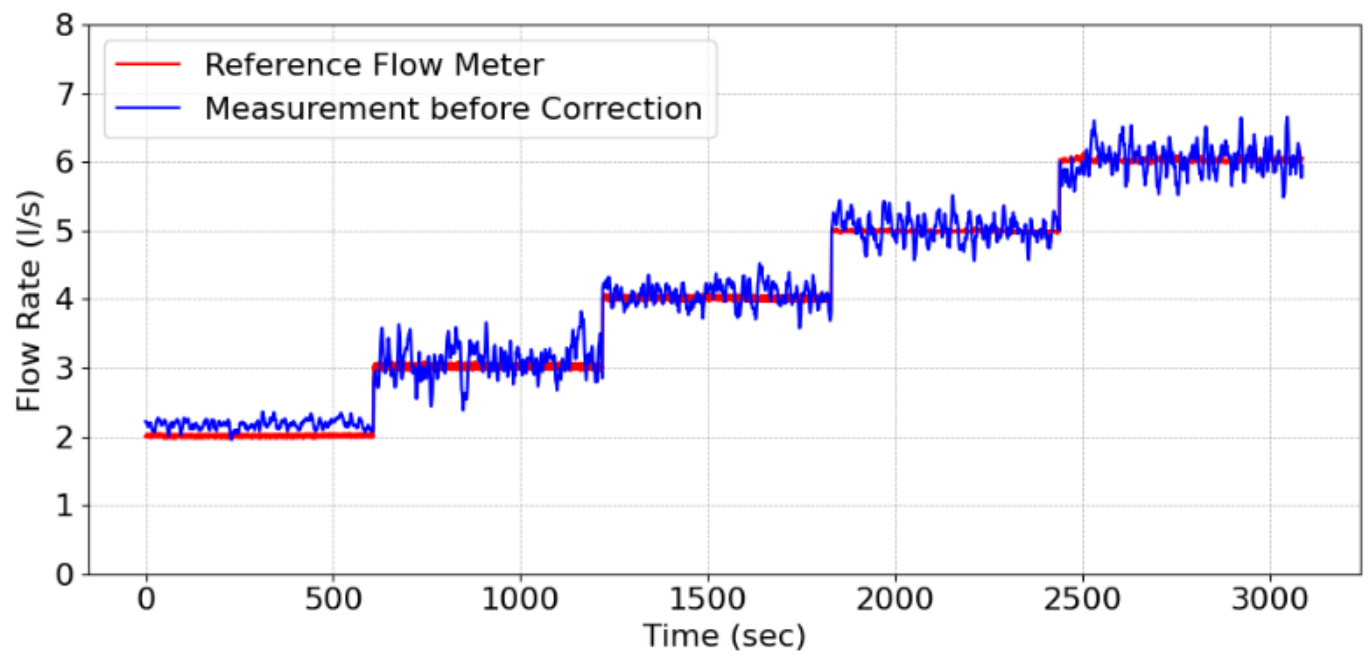

Fig. 6. Data from all measured flow rates (2 L/s to 6 L/s)

### B. Implementation of Flow Profile Correction Factor

The derived FPCF was applied to the measurement data to generate the corrected flow rates. Following the application of the FPCF, the measurement data become more consistent and suitable for calibration using a single, uniform calibration factor. This calibration factor is independent of variations in the velocity profile and is intended to compensate for other systematic influences, including sensor misalignment, installation effects, and environmental conditions. To evaluate the flow meter accuracy, the Flow-Weighted Mean Error (FWME) is determined as described in [27]. For this evaluation, six independent measurement trials, each lasting 100 seconds, were conducted at each specified flow rate. The first segment of each flow rate dataset was used to calibrate the flow meter, while the remaining five segments were employed to evaluate the flow meter's error after calibration. Table I and Fig. 7 compare the flow rate measurement errors before and after applying the FPCF and calibration. The results show that this combined procedure substantially improves accuracy, reducing the maximum error from 8.51% to 3.57% and lowering the FWME from 1.71% to 0.08%. These findings confirm the effectiveness of the proposed correction and calibration strategy across the tested flow range. It should be noted that the measurement accuracy reported in Table I represents the system-level performance, inherently encompassing the coupled contributions of the transducers, the timing and acquisition electronics, and the embedded signal-processing algorithms.

TABLE I
FLOW MEASUREMENT ERROR VS DIFFERENT FLOW RATES BEFORE AND AFTER FPCF & CALIBRATION

| Flow Rate (L/s) | Error (%) | |
|---|---|---|
| | Before FPCF & Calibration | After FPCF & Calibration |
| 2 | 8.51 | 3.57 |
| 3 | 2.62 | -0.42 |
| 4 | 1.46 | -0.44 |
| 5 | 0.42 | -0.60 |
| 6 | 0.23 | 0.07 |
| FWME | 1.71 | 0.08 |

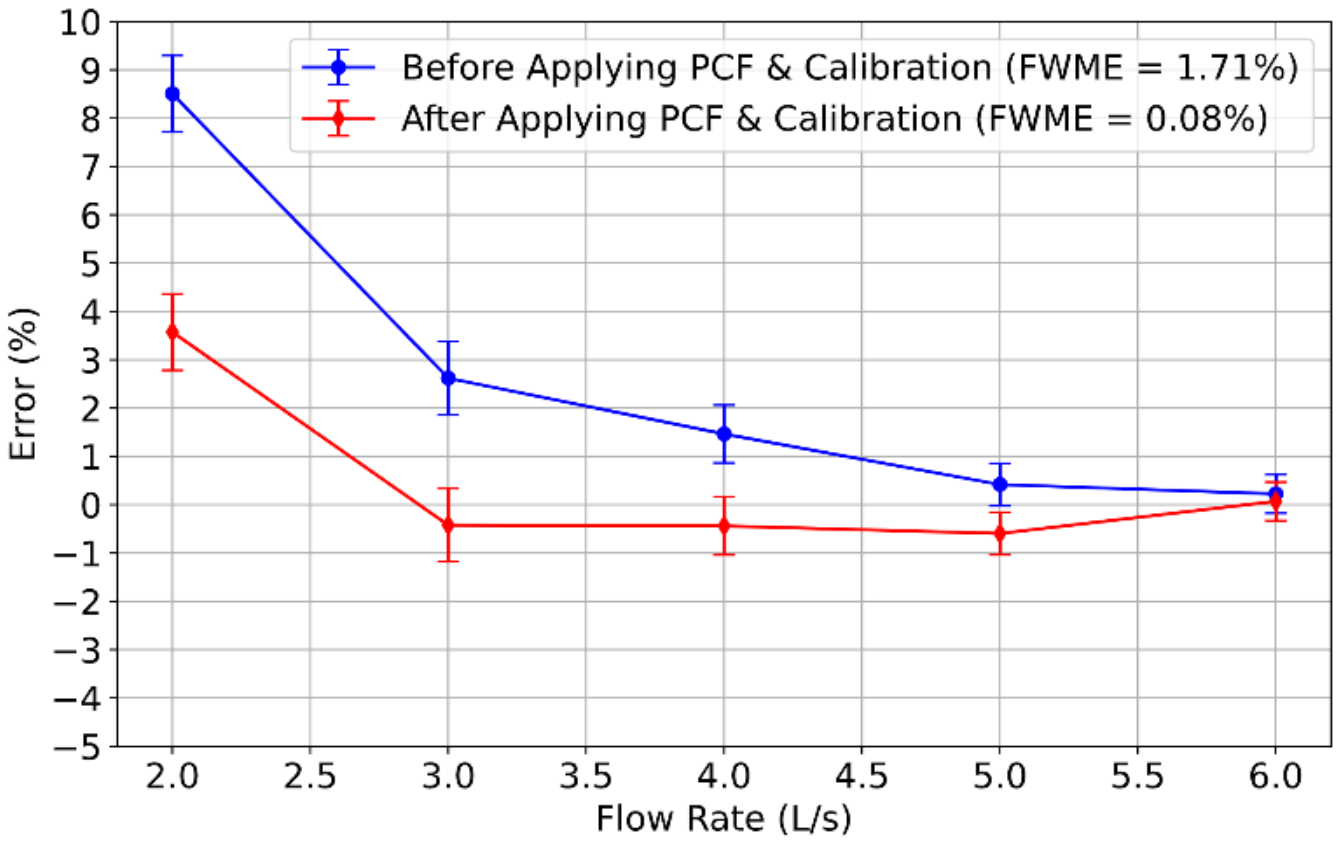

Fig. 7. Error with error bars for different flow rates before and after applying FPCF & calibration

### C. Repeatability Test of the Ultrasonic Flow Meter in Each Flow Rate

The repeatability of the ultrasonic flow meter was assessed through a series of controlled experiments at each flow rate, with a duration of 600 seconds for each. Repeatability was calculated using Equation (16), based on the standard deviation of $n$ repeated measurements, following the method described in [28]. The values are presented in Table II, which shows that the repeatability of the constructed flow meter is less than 0.8% at each flow rate, with an overall average of 0.6%.

$$R = \frac{\sigma}{\bar{Q}} \times 100\% = \frac{1}{\bar{Q}}\sqrt{\frac{1}{n-1}\sum_{i=1}^{n}(Q_i - \bar{Q})^2} \times 100\% \qquad (16)$$

where $Q_i$ is the i-th measured flow rate, $\bar{Q}$ is the mean flow rate, and $n$ is the number of repetitions.

TABLE II
FLOW MEASUREMENT ERROR VS DIFFERENT FLOW RATES AFTER APPLYING THE FPCF AND CALIBRATION

| Flow Rate (L/s) | Repeatability (%) |
|---|---|
| 2 | 0.79 |
| 3 | 0.76 |
| 4 | 0.60 |
| 5 | 0.44 |
| 6 | 0.40 |
| Average | 0.60 |

## D. Blockage Test and Clogging Detection

In real-world applications, partial blockages or flow obstructions may develop within the flow path, potentially disrupting normal operation. The timely detection of such anomalies and the ability to generate appropriate warnings are essential for ensuring system reliability and operational safety. To assess this diagnostic capability, an obstruction scenario was simulated by placing an acrylic weir directly upstream of the flow meter, resulting in upstream water accumulation and subsequent overflow. This experimental setup enabled evaluation of both the system's responsiveness to clogging conditions and the influence of partial blockage on key measurement parameters, including flow velocity, flow level, and overall accuracy. The results demonstrated that, despite the presence of partial obstruction, flow measurements remained viable, albeit with a modest increase in measurement error. In this test, two different weirs were used, as shown in Fig. 8. Weir 1 had a height equal to one-quarter of the pipe diameter, while Weir 2 had a height equal to one-half of the pipe diameter. The flow rate measurement error in this test is presented in Table III. Under these conditions, the error remains below 8% for all flow rates, indicating that flow measurement is still possible even in clogging or blockage conditions, although the error tends to increase.

Fig. 9 presents a comparative analysis of the experimental results obtained under three outlet configurations: No Weir, Weir 1, and Weir 2, across a range of flow rates. As anticipated, the introduction of a downstream weir leads to a reduction in flow velocity and an increase in fluid level. The relationship between velocity and level exhibits distinct patterns for each configuration, with measurable deviations from the baseline (No Weir) serving as indicators of partial obstructions within the pipe. This behavior forms the basis of an automatic obstruction detection mechanism: when a measured velocity–height data point falls significantly below the baseline curve, the system classifies it as an anomaly and generates a clogging alert. This self-diagnostic feature represents a novel capability of the developed flow meter, distinguishing it from most conventional systems that lack integrated blockage detection.

To establish the algorithmic logic for real-time clogging detection, we monitored the relationship between water height and average velocity. A decision boundary (threshold) was required to empirically separate the normal and clogging condition.

Following repeated experiments under both normal and partially clogged conditions, the following linear boundary was derived:

$$v = 0.00321H - 0.02 \tag{17}$$

where $H$(mm) is the water height and $v$(m/s) is the measured average velocity. Data points falling below this boundary are classified as belonging to the clogging zone. This boundary line was intentionally selected to lie between the data cluster corresponding to the normal-flow condition (No Weir) and the cluster corresponding to the initial stage of partial clogging condition (Weir 1).

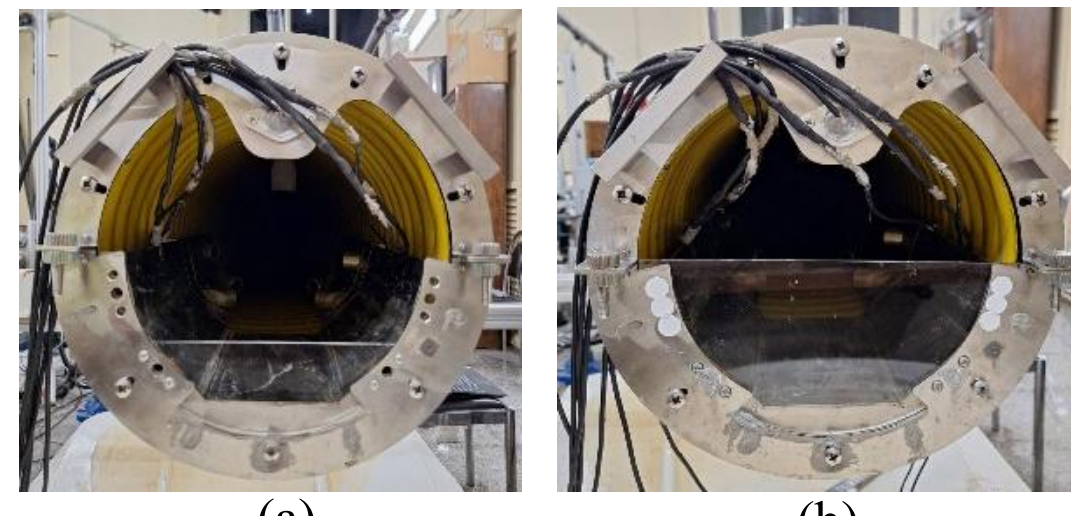

Fig. 8. Installation of the weirs in the ultrasonic flow meter for the blockage test; (a) Weir 1 one-quarter of the pipe diameter. (b) Weir 2 one-half of the pipe diameter.

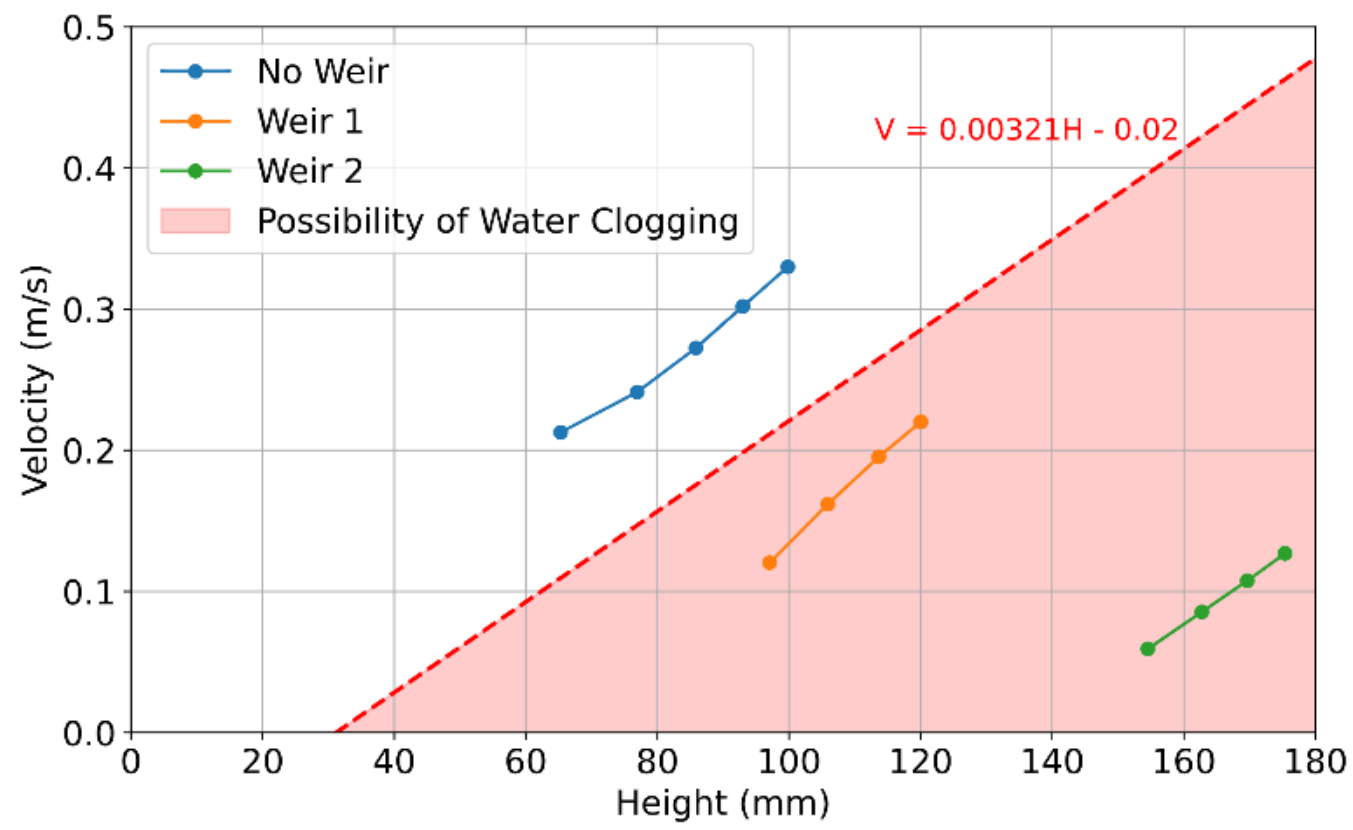

Fig. 9. Velocity vs. height in clogging detection test

TABLE III
FLOW MEASUREMENT ERROR VS DIFFERENT FLOW RATES IN BLOCKAGE TEST

| Flow Rate (L/s) | Error (%) | | |
|---|---|---|---|
| | No Weir | Weir 1 | Weir 2 |
| 2 | 3.57 | 3.31 | -2.57 |
| 3 | -0.42 | 7.61 | 0.33 |
| 4 | -0.44 | 7.15 | -0.81 |
| 5 | -0.60 | 4.95 | -1.85 |
| FWME | 0.08 | 5.91 | -1.19 |

## VI. CONCLUSION AND FUTURE WORKS

This study presents and experimentally validates an ultrasonic-based method for flow rate measurement in partially filled pipes. A custom-designed flow meter was developed to simultaneously measure velocity and water level, with the FPCF incorporated to account for non-uniform velocity distributions. Calibration and validation were performed using a dedicated flow loop with a high-precision Coriolis meter (±0.1% accuracy) as the reference. The experimental results demonstrate that the integrated application of FPCF and calibration markedly enhances measurement accuracy, decreasing the maximum error from 8.51% to 3.57% and reducing the FWME from 1.71% to 0.08%. These outcomes validate the effectiveness of the proposed correction–calibration framework in achieving reliable flow rate estimation across the investigated range. Furthermore, the system achieved measurement errors below 5% under baseline conditions and demonstrated diagnostic capability by detecting and alarming clogging scenarios in real time. In this study, validation was restricted to flow rates between 2 and 6 L/s. While the method remains applicable at higher Reynolds numbers, very high-speed flows with bubbles and strong surface fluctuations may reduce the accuracy of transit-time sensors. This limitation is acknowledged in our study. Future work will focus on integrating AI-based algorithms for adaptive correction and calibration, sensor configuration optimization using CFD, and the development of machine learning models to enhance robustness across diverse flow conditions.

## Acknowledgment

The authors gratefully acknowledge the partial support of this research by the Iran National Science Foundation (INSF) under Grant No. 4033633.

## References

[1] J. Chen, K. Zhang, L. Wang, and M. Yang, "Design of a high-precision ultrasonic gas flowmeter," *Sensors*, vol. 20, no. 17, p. 4804, 2020.

[2] L. Zhai, H. Xia, Y. Wu, and N. Jin, "Gas holdup measurement of horizontal gas–liquid two-phase flows using a combined ultrasonic–conductance sensor," *IEEE Sensors J.*, vol. 21, no. 24, pp. 27590–27600, 2021.

[3] H. C. H. Ng, H. L. Cregan, J. M. Dodds, R. J. Poole, and D. J. Dennis, "Partially filled pipes: Experiments in laminar and turbulent flow," *J. Fluid Mech.*, vol. 848, pp. 467–507, 2018.

[4] J. Brosda and M. Manhart, "Numerical investigation of semifilled-pipe flow," *J. Fluid Mech.*, vol. 932, A25, 2022.

[5] Y. Liu, T. Stoesser, and H. Fang, "Effect of secondary currents on the flow and turbulence in partially filled pipes," *J. Fluid Mech.*, vol. 938, A16, 2022.

[6] J. I. Yoon, J. Sung, and M. H. Lee, "Velocity profiles and friction coefficients in circular open channels," *J. Hydraulic Res.*, vol. 50, no. 3, pp. 304–311, 2012.

[7] A. Godley, "Flow measurement in partially filled closed conduits," *Flow Meas. Instrum.*, vol. 13, nos. 5–6, pp. 197–201, 2002.

[8] Y. Y. Qu, T. Lucke, and S. Beecham, "Measuring flows in partially-filled pipes in siphonic roof drainage systems," *Mapan*, vol. 26, no. 4, pp. 315–327, 2011.

[9] W. Li, Q. Zhang, X. Luo, and X. Chen, "New method of flow measurements based on CFD for partially filled pipe," in *Proc. Int. Conf. Mathematics, Modeling, Simulation and Statistics Application (MMSSA)*, 2018, pp. 49–53.

[10] X. Li and Z. Zhang, "Research on electromagnetic flowmeter measurement of non-full pipe flow based on numerical analysis method," in *Proc. IEEE ITNEC*, vol. 6, pp. 1326–1330, Feb. 2023.

[11] L. S. Nguyen *et al.*, "Vision-based system for the control and measurement of wastewater flow rate in sewer systems," *Water Sci. Technol.*, vol. 60, no. 9, pp. 2281–2289, 2009.

[12] H. W. Ji, S. S. Yoo, B. J. Lee, D. D. Koo, and J. H. Kang, "Measurement of wastewater discharge in sewer pipes using image analysis," *Water*, vol. 12, no. 6, p. 1771, 2020.

[13] G. Chen, G. Liu, B. Zhu, and W. Tan, "3D isosceles triangular ultrasonic path of transit-time ultrasonic flowmeter: Theoretical design and CFD simulations," *IEEE Sensors J.*, vol. 15, no. 9, pp. 4733–4742, 2015.

[14] N. Zhao, L. Hu, X. Peng, Z. Fang, W. Chen, and X. Fu, "A method combining measurement tool and numerical simulation for calculating acoustic signals of ultrasonic flowmeter," *IEEE Sensors J.*, vol. 19, no. 24, pp. 11805–11813, 2019.

[15] N. Munasinghe and G. Paul, "Ultrasonic-based sensor fusion approach to measure flow rate in partially filled pipes," *IEEE Sensors J.*, vol. 20, no. 11, pp. 6083–6090, 2020.

[16] M. Haide and W. Schroer, "Flow measurement in open channels using an ultrasonic phased array sensor," in *Proc. IEEE Sensors*, Nov. 2013, pp. 1–4.

[17] K. Klepiszewski, L. Solliec, and T. Bayer, "Flow monitoring under complex flow conditions using multiple sensors and COSP technology," in *Proc. IWA/IAHR Int. Conf. Urban Drainage*, 2017, pp. 10–15.

[18] M. Teufel and L. Solliec, "Using velocity profiles to determine an accurate volume flow rate at small and large dimensions," in *Proc. NOVATECH*, Lyon, France, 2010.

[19] M. Haide and W. Schroer, "An ultrasonic phased-array sensor to measure the velocity profile of open channels," in *Proc. OCEANS–San Diego*, Sept. 2013, pp. 1–4.

[20] H. Bonakdari and A. A. Zinatizadeh, "Influence of position and type of Doppler flow meters on flow-rate measurement in sewers using computational fluid dynamic," *Flow Meas. Instrum.*, vol. 22, no. 3, pp. 225–234, 2011.

[21] H. Bonakdari, "Establishment of relationship between mean and maximum velocities in narrow sewers," *J. Environ. Manage.*, vol. 113, pp. 474–480, 2012.

[22] M. Ronkin, A. Kalmykov, and K. Zeyde, "Novel FMCW-interferometry method testing on an ultrasonic clamp-on flowmeter," *IEEE Sensors J.*, vol. 20, no. 11, pp. 6029–6037, 2020.

[23] Y. Jiang, B. Li, and J. Chen, "Analysis of the velocity distribution in partially filled circular pipe employing the principle of maximum entropy," *PLoS One*, vol. 11, no. 3, p. e0151578, 2016.

[24] L. Peng, B. Zhang, H. Zhao, S. A. Stephane, H. Ishikawa, and K. Shimizu, "Data integration method for multipath ultrasonic flowmeter," *IEEE Sensors J.*, vol. 12, no. 9, pp. 2866–2874, 2012.

[25] H. Zhao, L. Peng, T. Takahashi, T. Hayashi, K. Shimizu, and T. Yamamoto, "ANN-based data integration for multi-path ultrasonic flowmeter," *IEEE Sensors J.*, vol. 14, no. 2, pp. 362–370, 2013.

[26] H. Zhao, L. Peng, T. Takahashi, T. Hayashi, K. Shimizu and T. Yamamoto, "CFD-Aided Investigation of Sound Path Position and Orientation for a Dual-Path Ultrasonic Flowmeter With Square Pipe," in *IEEE Sensors Journal*, vol. 15, no. 1, pp. 128-137, Jan. 2015, doi: 10.1109/JSEN.2014.2338322.

[27] ISO, *Measurement of fluid flow in closed conduits — Ultrasonic meters for gas — Part 1: Meters for custody transfer and allocation measurement*, ISO 17089-1:2019, 2nd ed., Aug. 2019.

[28] Joint Committee for Guides in Metrology (JCGM), *Evaluation of Measurement Data — Guide to the Expression of Uncertainty in Measurement*, JCGM 100:2008, 2008.